\documentclass{emulateapj}

\usepackage{graphicx}
\usepackage{epstopdf}

\slugcomment{Received 2009 November 14; accepted 2009 December 21}

\shorttitle{\emph{Kepler} Science Pipeline}
\shortauthors{Jenkins et al.}

\begin{document}

%% LaTeX will automatically break titles if they run longer than
%% one line. However, you may use \\ to force a line break if
%% you desire.

\title{Overview of the \emph{Kepler} Science Processing Pipeline}

%% Use \author, \affil, and the \and command to format
%% author and affiliation information.
%% Note that \email has replaced the old \authoremail command
%% from AASTeX v4.0. You can use \email to mark an email address
%% anywhere in the paper, not just in the front matter.
%% As in the title, use \\ to force line breaks.
\author{Jon M. Jenkins$^1$}%\altaffilmark{1} 
\author{Douglas A. Caldwell$^1$}%\altaffilmark{1}
\author{Hema Chandrasekaran$^1$}%\altaffilmark{1}
\author{Joseph D. Twicken$^1$}%\altaffilmark{1}
\author{Stephen T. Bryson$^2$}%\altaffilmark{3} 
\author{Elisa V. Quintana$^1$}%\altaffilmark{1}
\author{Bruce D. Clarke$^1$}%\altaffilmark{1}
\author{Jie Li$^1$}%\altaffilmark{1}
\author{Christopher Allen$^3$}%\altaffilmark{3}
\author{Peter Tenenbaum$^1$}%\altaffilmark{1}
\author{Hayley Wu$^1$}%\altaffilmark{1}
\author{Todd C. Klaus$^3$}%\altaffilmark{3}
\author{Christopher K. Middour$^3$}%\altaffilmark{3}
\author{Miles T. Cote$^3$}%\altaffilmark{3}
\author{Sean McCauliff$^3$}%\altaffilmark{3}
\author{Forrest R. Girouard$^3$}%\altaffilmark{3}
\author{Jay P. Gunter$^3$}%\altaffilmark{3}
\author{Bill Wohler$^3$}%\altaffilmark{3}
\author{Jeneen Sommers$^3$}%\altaffilmark{3}
\author{Jennifer R. Hall$^3$}%\altaffilmark{3}
\author{AKM K. Uddin$^3$}%\altaffilmark{3}
\author{Michael S. Wu$^5$}%\altafilmark{5}
\author{Paresh A. Bhavsar$^2$}%\altafilmark{5}
\author{Jeffrey Van Cleve$^1$}%\altaffilmark{1}
\author{David L. Pletcher$^2$}%\altaffilmark{1}
\author{Jessie A. Dotson$^2$}
\author{Michael R. Haas$^2$}
\author{Ronald L. Gilliland$^4$}%\altaffilmark{4}
\author{David G. Koch$^2$}%\altaffilmark{2}

%\and

\author{William J. Borucki$^2$}%\altaffilmark{3}

\email{Jon.Jenkins@nasa.gov}
\affil{$^1$SETI Institute/NASA Ames Research Center, M/S 244-30,  Moffett Field, CA 94035, USA}
\affil{$^2$NASA Ames Research Center, M/S 244-30, Moffett Field, CA 94035, USA}
\affil{$^3$Orbital Sciences Corporation/NASA Ames Research Center, M/S 244-30, Moffett Field, CA 94035, USA}
\affil{$^4$Space Telescope Science Institute, Baltimore, MD 21218, USA}
\affil{$^5$Bastion Technologies/NASA Ames Research Center, M/S 244-30, Moffett Field, CA 94035, USA}

%% Notice that each of these authors has alternate affiliations, which
%% are identified by the \altaffilmark after each name.  Specify alternate
%% affiliation information with \altaffiltext, with one command per each
%% affiliation.

%\altaffiltext{1}{Visiting Astronomer, Cerro Tololo Inter-American Observatory.
%CTIO is operated by AURA, Inc.\ under contract to the National Science
%Foundation.}
%\altaffiltext{2}{Society of Fellows, Harvard University.}
%\altaffiltext{3}{present address: Center for Astrophysics,
%    60 Garden Street, Cambridge, MA 02138}
%\altaffiltext{4}{Visiting Programmer, Space Telescope Science Institute}
%\altaffiltext{5}{Patron, Alonso's Bar and Grill}

%% Mark off your abstract in the ``abstract'' environment. In the manuscript
%% style, abstract will output a Received/Accepted line after the
%% title and affiliation information. No date will appear since the author
%% does not have this information. The dates will be filled in by the
%% editorial office after submission.

\begin{abstract}
The \emph{Kepler Mission} Science Operations Center (SOC) performs several critical functions including managing the $\sim$156,000 target stars, associated target tables, science data compression tables and parameters, as well as  processing the raw photometric data downlinked from the spacecraft each month. 
%The LC targets are sampled every 30 minutes and include all the planetary targets for which we seek signatures of transiting planets. The SC targets are sampled at 1-minute intervals permitting further characterization of the planet-star systems for the brighter ($Kp <$12) stars via asteroseismology, and more precise transit timing. 
The raw data are first calibrated at the pixel level to correct for bias, smear induced by a shutterless readout, and other detector and electronic effects. A background sky flux is estimated from $\sim$4500 pixels on each of the 84 CCD readout channels, and simple aperture photometry is performed on an optimal aperture for each star. Ancillary engineering data and diagnostic information extracted from the science data are used to remove systematic errors in the flux time series that are correlated with these data prior to searching for signatures of transiting planets with a wavelet-based, adaptive matched filter. Stars with signatures exceeding $7.1 \sigma$ are subjected to a suite of statistical tests including an examination of each star's centroid motion to reject false positives caused by background eclipsing binaries. Physical parameters for each planetary candidate are fitted to the transit signature, and signatures of additional transiting planets are sought in the residual light curve. The pipeline is operational, finding planetary signatures and providing robust eliminations of false positives.
\end{abstract}

%% Keywords should appear after the \end{abstract} command. The uncommented
%% example has been keyed in ApJ style. See the instructions to authors
%% for the journal to which you are submitting your paper to determine
%% what keyword punctuation is appropriate.


%% From the front matter, we move on to the body of the paper.
%% In the first two sections, notice the use of the natbib \citep
%% and \citet commands to identify citations.  The citations are
%% tied to the reference list via symbolic KEYs. The KEY corresponds
%% to the KEY in the \bibitem in the reference list below. We have
%% chosen the first three characters of the first author's name plus
%% the last two numeral of the year of publication as our KEY for
%% each reference.

%% Authors who wish to have the most important objects in their paper
%% linked in the electronic edition to a data center may do so by tagging
%% their objects with \objectname{} or \object{}.  Each macro takes the
%% object name as its required argument. The optional, square-bracket 
%% argument should be used in cases where the data center identification
%% differs from what is to be printed in the paper.  The text appearing 
%% in curly braces is what will appear in print in the published paper. 
%% If the object name is recognized by the data centers, it will be linked
%% in the electronic edition to the object data available at the data centers  
%%
%% Note that for sources with brackets in their names, e.g. [WEG2004] 14h-090,
%% the brackets must be escaped with backslashes when used in the first
%% square-bracket argument, for instance, \object[\[WEG2004\] 14h-090]{90}).
%%  Otherwise, LaTeX will issue an error. 

\section{Introduction}\label{s:intro}

The \emph{Kepler Mission} seeks to detect Earth-like planets transiting solar-like stars by performing photometric observations of $\sim$156,000 carefully selected target stars in \emph{Kepler's}  115 deg$^2$ field of view (FOV), as reviewed in \citet{borucki2010} and \citet{koch2010}. These Long Cadence (LC) targets are sampled every 29.4 minutes and include all the planetary targets for which we seek signatures of transiting planets. In addition, a total of 512 Short Cadence (SC) targets are sampled at 58.85 s intervals permitting further characterization of the planet-star systems for the brighter ($Kp <$12) stars via asteroseismology, and more precise transit timing. %\footnote{\emph{Kepler} is capable of processing 170,000 target definitions, but $\sim$13000 are used for calibration purposes or to monitor instrumental effects.}
 The \emph{Kepler Mission} Science Operations Center (SOC) at NASA Ames Research Center performs nine major functions:

\begin{enumerate}
\item Manage target aperture and definition tables specifying which  $5.4\times10^6$ of the $95\times 10^6$ pixels in the CCD array are processed and stored on the Solid State Recorder for later downlink\footnote{\emph{Kepler's} pointing stability requirement is 0".009, 3 $\sigma$, allowing us to preselect the pixels of interest for each star \citep{haas2010}.}.

\item Manage the science data compression tables and parameters, including the length-limited Huffman coding table, and the requantization table.

\item Report on the \emph{Kepler} photometer's health and status semiweekly after each X-band contact and monthly after each Ka-band science data downlink.

\item Monitor the pointing error and compute pointing tweaks when necessary to adjust the spacecraft pointing to ensure the validity of the uplinked science target tables, 

\item Process the science data each month to obtain calibrated pixels for all LC and SC targets, raw flux time series, and systematic error-corrected flux time series,\label{processingItem}.

\item Archive calibrated pixels, raw and corrected flux time series and centroid measurements to the Data Management Center (DMC).

\item  Search each flux time series for signatures of transiting planets,\label{transitSearchItem}.

\item Fit physical parameters and calculate error estimates for planetary signatures, and \label{ModelFitItem}.

\item Perform statistical tests to reject false positives %from sources such as background eclipsing binaries 
and establish accurate statistical confidence in each detection. \label{DataValidationItem}.
\end{enumerate}

%In this paper we concentrate on items \ref{processingItem}-\ref{DataValidationItem} but need to place the science processing in context with other supporting letters appearing in this issue.  \citet{haas2010} provides an overview of the \emph{Kepler Mission} from an operational perspective. \citet{batalha2010a} describes the selection of the ~155,000 planetary targets using stellar and observational characteristics to maximize the yield of planetary discoveries. The pixel-level calibrations performed at the SOC are informed by instrumental characteristics obtained during pre-flight testing and during the Science Commissioning period \citep{caldwell2010, vancleveandcaldwell2010}. The pixels of interest (POI) for each target star are determined by an optimal aperture algorithm detailed in \citet{bryson2010} that takes into account the Pixel Response Function (PRF), details of the CCD operation,  instrumental characteristics and the Kepler Input Catalog. This optimal aperture is used in the SOC Pipeline to extract simple aperture photometry and to construct stellar centroids for both use in spacecraft pointing management and for rejecting false positives due to background variables. Characteristics of the resulting LC flux time series are discussed in \citet{jenkins2010b} while those of the SC flux time series are discussed in \citet{gilliland2010b}. The process of vetting transit candidates using both \emph{Kepler} data and Follow Up Observation (FOP) data is described in \citet{batalha2010b} while the FOP Program is discussed in \citet{gautier2010}.  
 
\emph{Kepler's} observations are organized into three month intervals called quarters defined by the roll maneuvers the spacecraft executes about its boresight to keep the solar arrays pointed toward the Sun  \citep{haas2010}. Once each month,  the accumulated science data are transmitted via the Deep Space Network\footnote{The DSN is operated by the Jet Propulsion Laboratory for NASA.}  (DSN) to the Mission Operations Center\footnote{The MOC is located at LASP in Boulder, CO, USA.}, which  forwards them to the DMC\footnote{The DMC is located at the Space Telescope Science Institute (STScI) in Baltimore, MD, USA.}. The DMC packages them into FITS files and pushes them to the SOC. A selected set of ancillary engineering data are also delivered with the science data, containing any parameters likely to have a bearing on the quality of the science data, such as temperature measurements of the focal plane and readout electronics. 

\begin{figure}
%\epsscale{.80}
\plotone{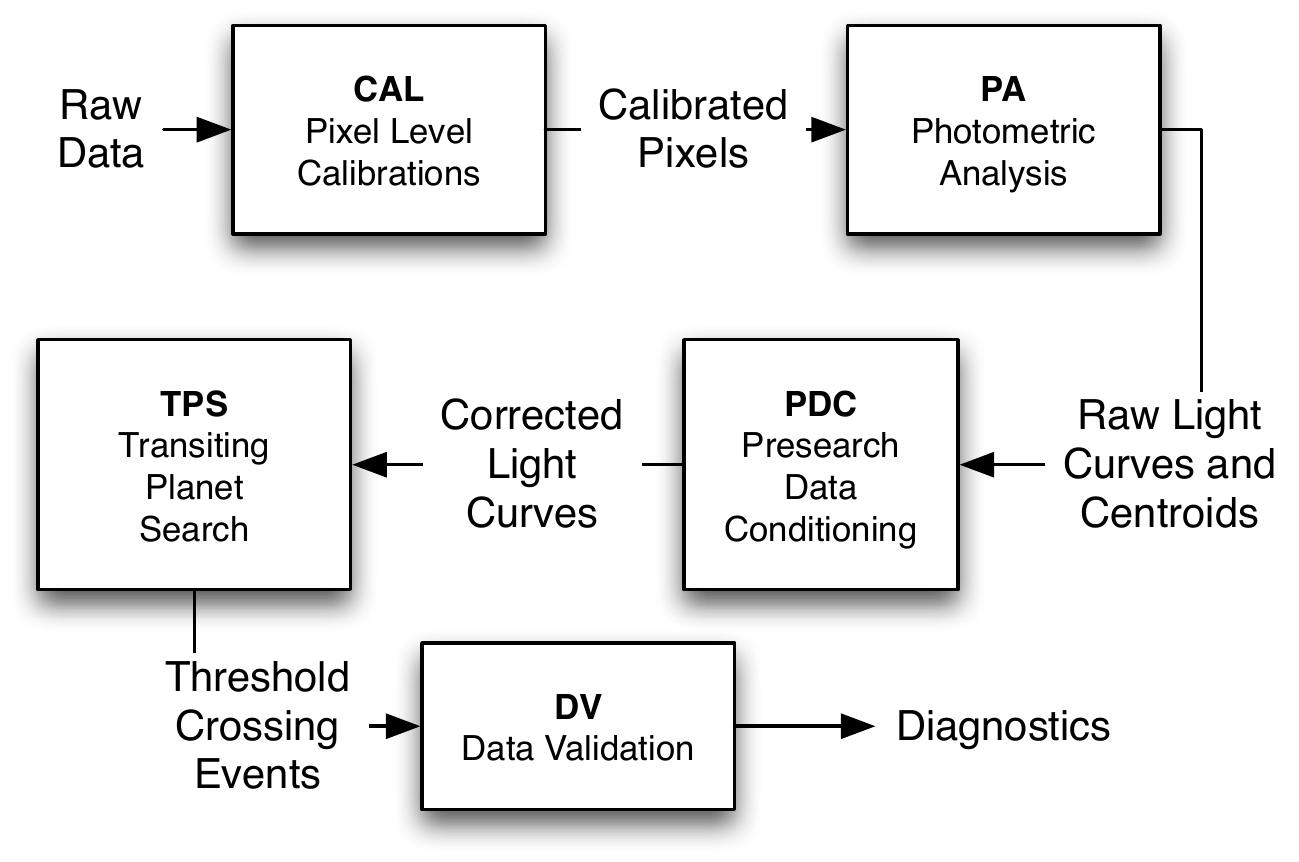}%SOC_SciencePipeline.pdf}
\caption{Data flow diagram for the SOC Science Pipeline. 
\label{fig:SOC_Pipeline}}
\end{figure}

The Science Pipeline is divided into several components in order to allow for efficient management and parallel processing of the data, as shown in Figure \ref{fig:SOC_Pipeline}.  Raw pixel data downlinked from the \emph{Kepler} photometer are calibrated by the Calibration module (CAL) to produce calibrated target and background pixels and their associated uncertainties. The calibrated pixels are processed by Photometric Analysis (PA) to fit and remove sky background and extract simple aperture photometry from the background-corrected, calibrated target pixels. PA also measures the centroid locations of each star on each frame. The final step to produce light curves happens in Pre-search Data Conditioning (PDC) where signatures in the light curves correlated with systematic error sources such as pointing drift, focus changes, and thermal transients are removed. Output data products include raw and calibrated pixels, raw and systematic error-corrected flux time series, centroids and associated uncertainties for each target star, which are archived to the DMC and eventually made available to the public through the Multimission Archive at STScI.\footnote{http://stdatu.stsci.edu/kepler/}

In Transiting Planet Search (TPS) a wavelet-based, adaptive matched filter is applied to identify transit-like features with durations in the range of 1--16 hours. Light curves with transit-like features whose combined (folded) transit detection statistic exceeds 7.1$\sigma$ for some trial period and epoch are designated as Threshold Crossing Events (TCEs) and subjected to further scrutiny by Data Validation (DV). This threshold ensures that no more than one false positive will occur due to random fluctuations over the course of the mission, assuming non-white, non-stationary Gaussian observation noise \citep{jenkinsCaldwellAndBorucki2002,jenkins2002}. DV performs a suite of statistical tests to evaluate the confidence in the detection, to reject false positives by background eclipsing binaries, and to extract physical parameters of each system (along with associated uncertainties and covariance matrices) for each planet candidate. After the planetary signature has been fitted, it is removed from the light curve and the residual is subjected to a search for additional transiting planets. This process repeats until no further TCEs are identified. The DV results and diagnostics are furnished to the Science Team to facilitate disposition by the Follow-up Observing Program \citep[FOP;][]{gautier2010}.

\section{Pixel Level Calibrations}\label{s:CAL}
\begin{figure}
%\epsscale{.80}
\plotone{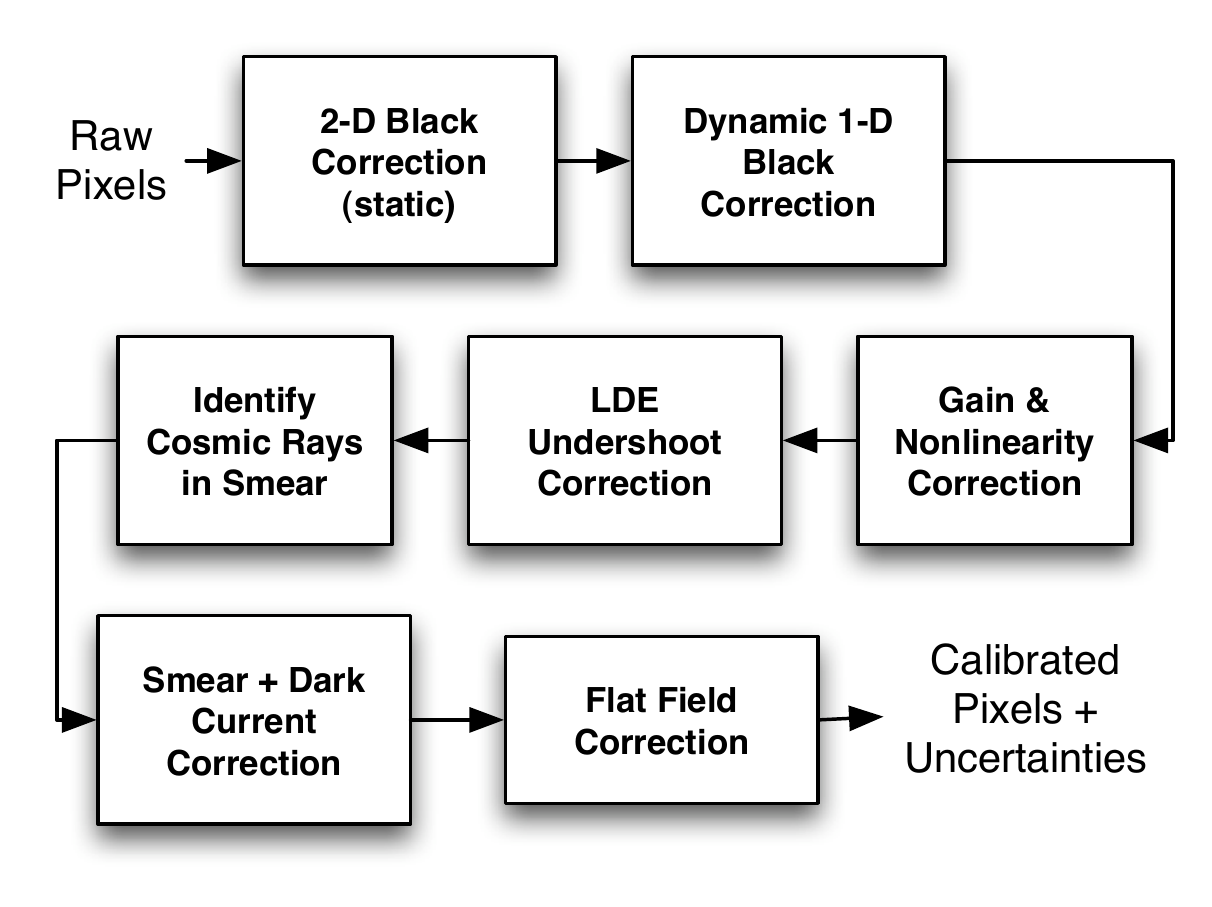}%CAL.pdf}
\caption{Data flow diagram for the Calibration Pipeline module. \label{fig:CAL}}
\end{figure}

The Pipeline module CAL corrects the raw \emph{Kepler} photometric data at the pixel level prior to the extraction of photometry and astrometry. Several of the processing steps given in  Figure \ref{fig:CAL} are familiar to ground-based photometrists. However, a few are peculiar to \emph{Kepler} due to the lack of a shutter and unique features in its analog electronics chains. Details of these instrument characteristics and how they were determined and updated in flight are discussed in \citet{caldwell2010} and are comprehensively documented in the \emph{Kepler} Instrument Handbook \citep{vancleveandcaldwell2010}. 

The sequence of processing steps  in CAL that produce calibrated pixels and associated uncertainties is as follows. (1)  The two-dimensional black level (CCD bias voltage) structure (fixed pattern noise) is removed, followed by fitting and removing a dynamic estimate of the black level. (2) Gain and nonlinearity corrections are applied. (3) The analog electronics chain exhibits memory, necessitating the application of a digital filter to remove this effect, called Local Detector Electronics (LDE) undershoot. (4) Cosmic ray events in the black and smear measurements are removed prior to subsequent corrections. (5) The smear signal caused by operating in the absence of a shutter and the dark current for each CCD readout channel are estimated from the masked and virtual smear collateral data measurements. (6) A  flat field correction is applied. 

%Most of the calibration models were determined before launch during preflight testing and characterization, although some were updated during commissioning (e.g., we obtained dark frames prior to releasing the dust cover on April 7 2009), and some are updated through occasional non-science mode data collections (such as reverse-clocked images), and some are monitored via witness pixels simultaneously with the science data (such as the two-dimensional bias voltage structure).

%\begin{figure}
%%\epsscale{.80}
%\plotone{f2.eps}%CAL.pdf}
%\caption{Data flow diagram for the Calibration Pipeline module. \label{fig:CAL}}
%\end{figure}

\section{Photometric Analysis}\label{s:PA}
Before photometry and astrometry can be extracted from the calibrated pixel time series, the Pipeline detects so-called "Argabrightening" events in the background pixel data. These mysterious transient increases in the background flux were identified early in  Commissioning. The current hypothesis is that these transient events are due to small dust particles from \emph{Kepler} achieving escape velocity after micrometeorite hits and reflecting sunlight into the barrel of the telescope as they drift across the FOV. Argabrightenings that affect 10 or more CCD readout channels occur $\sim$15 times per month, but the rate is dropping over time. The most egregious of these events cannot be perfectly corrected by the current background correction. We gap the data when the excess background flux exceeds the 100 Median Absolute Deviation (MAD) level. 

PA then robustly fits a two-dimensional surface to $\sim$4500 background pixels on each channel  to estimate the sky background, which is evaluated at each target star pixel location and subtracted from the calibrated pixel values. %Cosmic ray events are also identified in the calibrated background pixel time series and flagged, but since the sky background uses a robust fit, the resulting correction is not sensitive to the effectiveness of the cosmic ray identification.
Each target pixel time series is scanned for cosmic rays by first detrending the time series with a moving median filter with a width of five cadences (time steps) and examining the residuals for outliers compared to the MAD of the residuals for each pixel. Care is taken not to remove clusters of outliers that might be due to astrophysical signatures such as flares or transits that are intrinsic to the target star. 

The photocenters of the 200 brightest, unsaturated stars on each channel are fitted using the pixel response functions \citep[PRF;][]{bryson2010} and then used to define the ensemble motion of the stars over the observations. The aggregate star motion is used along with the PRFs reconstructed from Commissioning data to define the optimal aperture as the collection of pixels that maximizes the mean signal-to-noise ratio of the flux measurement for each star \citep{bryson2010}. The background-corrected, cosmic ray-corrected pixels are then summed over the optimal aperture to define a flux estimate for each cadence frame. 

%Future enhancements to the PA Pipeline include the capability to perform Difference Image Analysis photometry similar to that described in \citet{gilliland2000}.

\section{Systematic Error Corrections}\label{s:PDC}
PDC's task is to remove systematic errors from the raw flux time series. These include pointing errors, focus changes, and thermal effects on instrument properties. PDC co-trends each flux time series against ancillary engineering data such as temperatures of the focal plane and electronics, reconstructed pointing and focus variations to remove signatures correlated with these proxy systematic error measurements. 
A Singular Value Decomposition (SVD) is applied to the design matrix containing the ancillary data to identify the most significant, independent components and to stabilize the matrix inversion inherent in the fit to the data. 
Additionally, PDC identifies residual isolated outliers, and fills intra-quarter gaps so that the data for each quarterly segment are contiguous when presented to TPS. Finally, PDC adjusts the light curves to account for the excess flux in the optimal apertures due to starfield crowding in order to make apparent transit depths uniform from quarter to quarter as the stars move from detector to detector with each roll maneuver. This is achieved by estimating the mean excess flux in each photometric aperture from sources other than the target star itself from knowledge of the PRF and background star positions and magnitudes and subtracting this value from each point in the time series  \citep[see ][]{bryson2010}.
%An examples of raw and corrected flux time series from the first two months of Q2 are presented in Figure \ref{fig:PDC}.

Significant effort has been applied to PDC in order to achieve good results with flight data. There are a number of phenomena that were significantly different than expected, including focus variations, and the amount of pointing drift observed during the first two quarters of operation. 
%%Figure \ref{fig:Mod2Out1PlatescaleAndDrift} shows the average centroid drift of stars on Module 2 Output 1 near the outside corner of the focal plane array (FPA) along with the change in the platescale observed during quarter 1 (Q1). 
%The pointing drift has slowed significantly now that we've switched out several variable stars on the Fine Guidance Sensors (FGS) whose intensity varations were modulating the centroid information fed to the spacecraft's Attitude Determination and Control System (ADCS). Some software parameters for the FGS centroiding algorithm have also been tuned to make the centroiding less vulnerable to variable stars and time-varable sky background. 
The systematic errors observed in flight exhibit a range of different time scales, from a few hours to several days to many days and weeks. Such phenomena include the intermittent modulation of the focus by $\sim$1 $\mu$m every 3.2 hr by a heater on one of the reaction wheel assemblies. One of the Fine Guidance Sensors' guide stars through the first quarter (Q1) of observations was an eclipsing binary whose 30$\%$ eclipses induced a 1 mpix pointing excursion lasting $\sim$8 hr every 1.7 days \citep{haas2010}. By far the strongest systematic effects in the data so far have occurred after each of two safe mode events \citep{haas2010} during which the photometer was shut off, the telescope cooled and the focus changed by $\sim$2.2 $\mu$m per $^\circ$C. One of these occurred at the end of Q1 and the second $\sim$2 weeks into Q2. Thermal effects can be observed in the science data for $\sim$5 days after each safe mode recovery. The fact that most systematics such as these affect all the science data simultaneously, and that there is a rich amount of ancillary engineering data and science diagnostics available provides significant leverage in dealing with these effects. 
%We have found, however, that most such ancillary information themselves are superpositions of the various effects, and that a modification of PDC to apply the co-trending to band-pass filtered versions of the flux time series and the ancillary data has demonstrated good results. 

Some systematic phenomena are specific to individual stars and cannot be corrected by co-trending against ancillary data. The first issue is the occasional, abrupt drop in pixel sensitivity that introduces a step discontinuity in an affected star's light curve (and associated centroids). This is often preceded immediately by a cosmic ray event, and is sometimes followed by an exponential recovery over a few hours, but usually not to the same flux level as before. The typical drop in sensitivity is 1$\%$, which is unmistakable in the flux time series.  Such step discontinuities are identified separately from those due to operational activities, such as safe modes and pointing tweaks, and are mended by raising the light curve after the discontinuity for the remainder of the quarter. These events do not mimic transits since they do not recover to the same pre-event flux level, and few transits, if any, are affected by this correction. The second issue is that many stars exhibit coherent or slowly evolving oscillations that interfere with systematic error removal. The approach taken is to identify and remove strong coherent components in the frequency domain prior to co-trending, and then to restore these components to the residuals after co-trending.

\begin{figure}
%\epsscale{.80}
\plotone{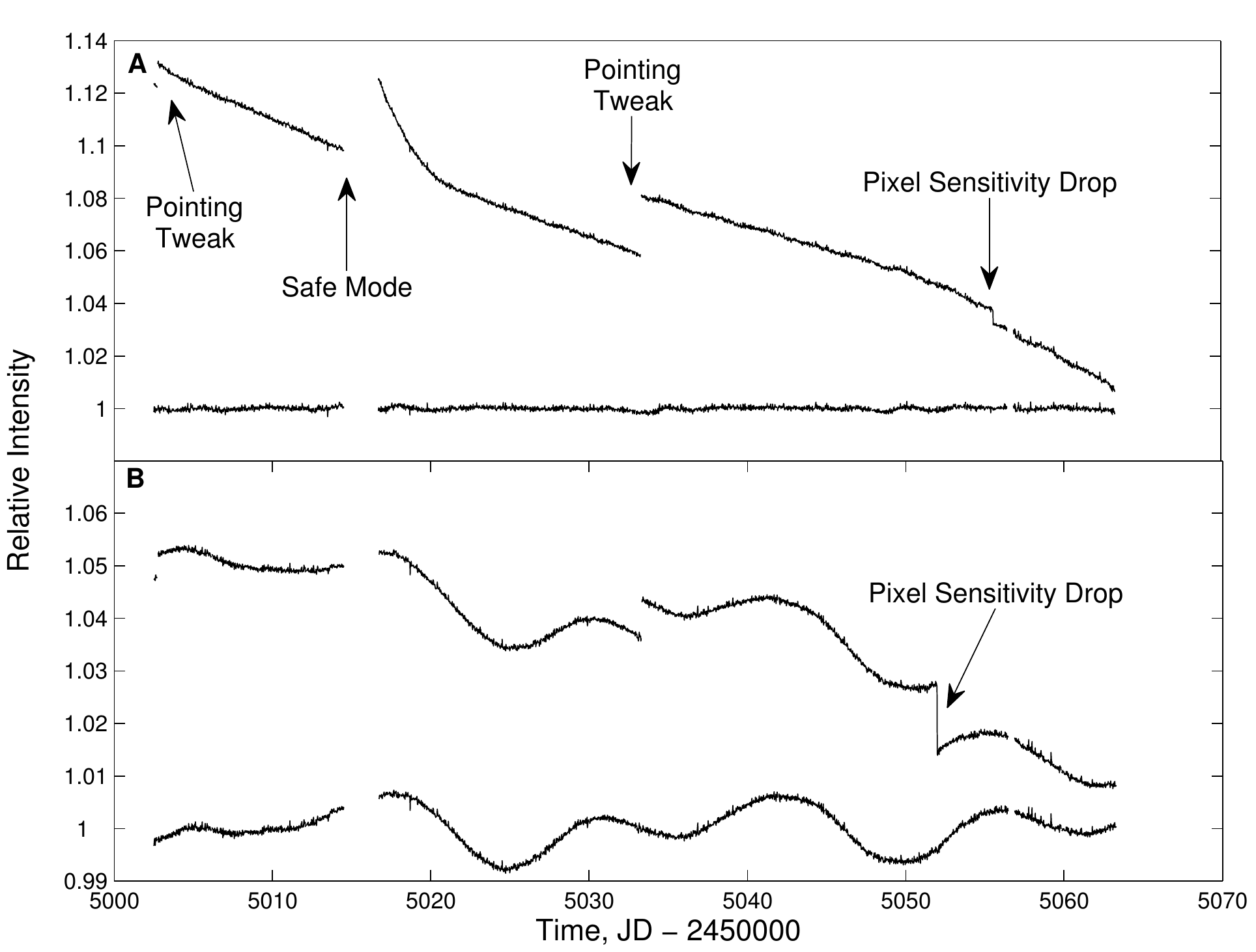}%pdcFluxVsTimeTwoStars.pdf}
\caption{Raw and systematic error-corrected light curves for two different stars. The raw light curves exhibit step discontinuities due to pointing offsets, thermal transients due to safe modes and pixel sensitivity changes, as well as pointing errors and focus changes, as indicated in the figure. Panel A shows the raw flux time series (top curve, offset) and the PDC flux time series (bottom curve) of a $Kp=15.6$ dwarf star. Panel B shows the raw and PDC flux time series for a $Kp=14.9$ dwarf star that displays less sensitivity to the thermal transients. Both stars' corrected flux time series are significantly improved by the detrending afforded by PDC while retaining intrinsic stellar variability on timescales of several weeks.
\label{fig:PDC}}
\end{figure}

Figure \ref{fig:PDC} shows the results of running two flux time series obtained during Quarter 2 through PDC on schedule for release early in 2010, demonstrating PDC's effectiveness. We expect that learning to deal with the various systematic errors will consume a great deal of effort over the lifetime of the mission as we push the detection limit to smaller and smaller planets.

%\begin{figure}
%%\epsscale{.80}
%\plotone{f3.eps}%pdcFluxVsTimeTwoStars.pdf}
%\caption{The results of running two different light curves through Pre-search Data Conditioning (PDC), which has the ability to remove step discontinuities due to pointing offsets, thermal transients due to safe modes and pixel sensitivity changes, as well as pointing errors and focus changes, as indicated in the figure. Panel A shows the raw flux time series (top curve, offset) and the PDC flux time series (bottom curve) of a $Kp=15.6$ dwarf star. Panel B shows the raw and PDC flux time series for a $Kp=14.9$ dwarf star that displays less sensitivity to the thermal transients. Both targets' corrected flux time series are significantly improved by the detrending afforded by PDC while retaining intrinsic stellar variability on timescales of several weeks.
%\label{fig:PDC}}
%\end{figure}
%%\clearpage

\section{Transiting Planet Search}\label{s:TPS}
TPS searches for transiting planets by "stitching" the quarterly segments of data together to remove gaps and edge effects and then applies the wavelet-based, adaptive matched filter of \citet{jenkins2002}. This approach is a time-adaptive approach that estimates the power spectrum of the observational noise as a function of time% in a natural, smooth fashion
. This approach was developed specifically for solar-like stars with colored, broadband power spectra. Some modifications to the original approach have been developed to accommodate target stars that exhibit coherent structure in the frequency domain. Similar to the approach adopted in PDC, we fit and remove strong harmonics that are  inconsistent with transit signatures prior to applying the wavelet-based filter. This significantly increases the sensitivity of the transit search for such stars and also provides photometric precision estimates (as by-products of the search) that are more realistic for such targets. If the transit-like signature of a given target star exceeds 7.1 $\sigma$ then a TCE is recorded and sent to DV for additional scrutiny.

\section{Data Validation}\label{s:DV}

DV performs a suite of tests to establish or break confidence in each TCE flagged by TPS, as well as to fit physical parameters to each transit-like signature. DV is currently under development and we anticipate its release in early 2010 to support the next FOP observing season.

The statistical confidence in the TCE is examined by performing a bootstrap test \citep{jenkins2002, jenkinsCaldwellAndBorucki2002} to take into account non-Gaussian statistics of the individual light curves. A transiting planet model is fitted to the transit signature as a joint noise characterization/parameter estimation problem. That is, the observation noise is \emph{not} assumed to be white and its characteristics are estimated using the wavelet-based approach employed in TPS, but as an estimator, rather than as a detector. This process yields a set of physical parameters and an associated covariance matrix.

To eliminate false positives due to eclipsing binaries, the planet model fit is performed again only to the even transits, and then only to the odd transits, and the resulting odd/even depths and epochs are compared in order to see if the results indicate the presence of secondary eclipses. After the multi-transiting planet search is complete, the periods are compared to detect eclipsing binaries with significant eccentricity causing TPS to detect two transit pulse trains at essentially the same period, but at a phase other than 0.5.

To guard against background eclipsing binaries, a centroid motion test is performed to determine whether the centroids shifted during the transit sequence. If so, the source right ascension and declination can be estimated by the measured in- versus out-of-transit centroid shift normalized by the fractional change in brightness of the system \citep[i.e., the transit "depth";][]{batalha2010,monet2010}.

Additional tests include checking whether the transit signature is consistent in the target pixels, whether the transit signature is correlated with any ancillary engineering data or any collateral data, and whether the distribution of cosmic ray events during transit is significantly different than that out-of-transit.

\section{Future Development}\label{s:futureDevelopment}
Future development for the SOC includes implementing difference image analysis photometry, completing DV, and developing and implementing mitigations for the instrument artifacts described in \citet{caldwell2010}. These artifacts affect a small portion of the \emph{Kepler} FOV at any one time. The development schedule calls for delivery of these features to allow us to discover long period Earths by late 2010 and recover at least 90\% of the FOV  in 2011 in time to  characterize the frequency of Earth-size transiting planets in the habitable zones of solar-like stars.

The instrumental artifacts consist of two categories of phenomena:  (1) temperature-dependent two-dimensional bias image structure, and other temperature-dependent electronics effects, and (2) Moir\'e patterns caused by an unstable circuit with an operational amplifier oscillating at  $\sim$1.5 GHz. Normally this latter feature appears as a high-frequency oscillation on each readout row whose frequency changes with row number as the readout electronics heat up during readout. When this signal aliases to the sample rate of the CCD readout, a transient band appears in an affected channel and slowly rolls across the frame as the temperature changes. The Moir\'e pattern can interact with bright, saturated star signals and generate scene-dependent effects. The typical amplitude of these image artifacts is $<$1 ADU per pixel per read, comparable to or smaller than the typical readout noise. It is important to note that not all of these op amps are oscillating and that the perturbations to the images are very small. Our mitigation plan consists of two approaches, one for the temperature-dependent effects, and one for the Moir\'e pattern effects, and most of the effort takes place prior to pixel level calibrations. 

Before launch, we added pixels to the target table that allow us to sample and trend the artifacts simultaneously with the science data. The \emph{Kepler} Science Office and SOC are developing and prototyping algorithms that use these image artifact pixels, together with other science data, to reconstruct the underlying temperature-dependent two-dimensional bias structure as a function of time over each quarter. The resulting dynamic model will allow the temperature-dependent bias signals to be removed directly from the data. Moreover,  the thermal environment of the \emph{Kepler} photometer is very stable and changes slowly during nominal operations. Thus any residual thermal two-dimensional bias effects will be small after the corrections are in place and can be co-trended out of the data by PDC like other thermally driven, instrumental effects. 

Given that the Moir\'e pattern noise exhibits both high spatial frequencies and high temporal frequencies, the prospect of reconstructing a high fidelity model of the effects at the pixel level with an accuracy sufficient to \emph{correct} the affected data appears unlikely. We are developing algorithms that identify when these Moir\'e patterns are present and mark the affected CCD regions as suspect on each affected LC. These suspect data flags will then be used to inform downstream modules so that the affected data can be appropriately weighted, and so that, for example, TPS can selectively ignore time intervals that are potentially contaminated with electronics-induced transients when searching for transit signatures. DV will produce a contamination report for each TCE indicating the fraction and severity of the Moir\'e pattern effect. The Pipeline will track and trend diagnostic metrics reflecting the prevalence and severity of the Moir\'e pattern as a diagnostic of this aspect of the photometer performance.

In spite of the presence of these image artifacts, \emph{Kepler} is already achieving photometric precision  sufficient to detect Earth-size planets transiting solar-like stars for the majority of the FOV at any given time \citep{jenkins2010}. We are confident that these efforts will enable us to minimize the impact of these artifacts on exoplanet detection, and produce high-quality photometric and astrometric time series for other scientific investigations by the greater astronomical community.

\section{Conclusions}\label{s:Conclusions}
We have presented an overview of the \emph{Kepler} SOC science pipeline processing. The output products include raw and calibrated pixel time series, raw and systematic error-corrected flux time series, centroid time series for each star, and associated uncertainties. These products will permit the detection and characterization of transiting planets in the \emph{Kepler} FOV as well as enabling astrophysical investigations and serendipitous discoveries not contemplated in \emph{Kepler's} driving design requirements.

\acknowledgments

Funding for this Discovery Mission is provided by NASA's Science Mission Directorate.
We thank thousands of people whose efforts made \emph{Kepler's} grand voyage of discovery possible.

\clearpage

%% Use the figure environment and \ one or \plottwo to include
%% figures and captions in your electronic submission.
%% To embed the sample graphics in
%% the file, uncomment the \plotone, \plottwo, and
%% \includegraphics commands
%%
%% If you need a layout that cannot be achieved with \plotone or
%% \plottwo, you can invoke the graphicx package directly with the
%% \includegraphics command or use \plotfiddle. For more information,
%% please see the tutorial on "Using Electronic Art with AASTeX" in the
%% documentation section at the AASTeX Web site,
%% http://www.journals.uchicago.edu/AAS/AASTeX.
%%
%% The examples below also include sample markup for submission of
%% supplemental electronic materials. As always, be sure to check
%% the instructions to authors for the journal you are submitting to
%% for specific submissions guidelines as they vary from
%% journal to journal.

%% This example uses \plotone to include an EPS file scaled to
%% 80% of its natural size with \epsscale. Its caption
%% has been written to indicate that additional figure parts will be
%% available in the electronic journal.

\clearpage

\clearpage

\clearpage

\clearpage

\clearpage

%% The following command ends your manuscript. LaTeX will ignore any text
%% that appears after it.

\end{document}